\documentclass[reprint, superscriptaddress, amsmath, amssymb, aps,
pra]{revtex4-2}
\usepackage{graphicx}
\usepackage{dcolumn}
\usepackage{bm}
\usepackage{subfigure}
\usepackage{hyperref}

\begin{document}

\title{Chiral Transfer and Entanglement Generation of Even-Parity Bell States
with Engineered Two-Photon Loss}

\author{Lin Xiao}
\affiliation{School of Electronic Science and Engineering, Chongqing University
of Posts and Telecommunications, Chongqing 400065, China}
\author{Jian Li}
\affiliation{School of Electronic Science and Engineering, Chongqing University
of Posts and Telecommunications, Chongqing 400065, China}
\affiliation{Chongqing Key Laboratory of Dedicated Quantum Computing and Quantum
Artificial Intelligence, Chongqing, 400065, China}
\affiliation{Southwest Center for Theoretical Physics, Chongqing University,
Chongqing 401331, China}
\author{Mu Zhou}
\affiliation{School of Electronic Science and Engineering, Chongqing University
of Posts and Telecommunications, Chongqing 400065, China}
\affiliation{Chongqing Key Laboratory of Dedicated Quantum Computing and Quantum
Artificial Intelligence, Chongqing, 400065, China}
\author{Bin Wei}
\email{weibin@cqu.edu.cn}
\affiliation{Department of Physics and Chongqing Key Laboratory for Strongly
Coupled Physics, Chongqing University, Chongqing 401331, China}
\author{Qing-Xu Li}
\email{liqx@cqupt.edu.cn}
\affiliation{School of Electronic Science and Engineering, Chongqing University
of Posts and Telecommunications, Chongqing 400065, China}
\affiliation{Chongqing Key Laboratory of Dedicated Quantum Computing and Quantum
Artificial Intelligence, Chongqing, 400065, China}
\author{Jia-Ji Zhu}
\email{zhujj@cqupt.edu.cn}
\affiliation{School of Electronic Science and Engineering, Chongqing University
of Posts and Telecommunications, Chongqing 400065, China}
\affiliation{Chongqing Key Laboratory of Dedicated Quantum Computing and Quantum
Artificial Intelligence, Chongqing, 400065, China}
\affiliation{Southwest Center for Theoretical Physics, Chongqing University,
Chongqing 401331, China}


\begin{abstract}
	We investigate chiral transfer and dissipative generation of even-parity Bell
states in a two-qubit system with coherent two-photon driving and engineered
two-photon loss. We find that adiabatic encirclement of a second-order
exceptional point induces direction-dependent transfer between the Bell states
$|\Phi^+\rangle$ and $|\Phi^-\rangle$, governed by a time-integrated low-loss
branch-selection mechanism. We develop a hybrid-Liouvillian description with a
control parameter $q$ that interpolates between conditional no-jump dynamics and
unconditional Lindblad evolution, and use it to assess how
exceptional-point-induced chirality survives in the presence of quantum jumps.
When the same parameter loop is initialized in the separable state $|00\rangle$,
it directly generates strong even-parity entanglement. Together, these results
extend dissipative Bell-state control beyond the single-excitation manifold.
They further demonstrate that exceptional-point-based protocols can unify chiral
state transfer and entanglement generation within a single engineered two-photon
platform.
\end{abstract}

\maketitle

\section{INTRODUCTION}

Control of quantum states in open systems has traditionally focused on
suppressing environmental coupling in order to preserve
coherence~\cite{nakamuraCoherentControlMacroscopic1999,%
liCommunicationsSpatialSeparation2010,%
jingzhangProtectingCoherenceEntanglement2010,%
altafiniCoherentControlOpen2004,%
zhangAsymptoticallyNoiseDecoupling2007,%
gordonUniversalDynamicalDecoherence2007,%
manCavitybasedArchitecturePreserve2015,%
zurekDecoherenceEinselectionQuantum2003b,%
violaDynamicalDecouplingOpen1999a,%
lidarDecoherenceFreeSubspacesQuantum1998a,%
yangPreservingQubitCoherence2011}.
Recent advancements in non-Hermitian physics and reservoir engineering have
demonstrated that dissipation can be harnessed as a resource for quantum-state
preparation and control~\cite{jinBulkboundaryCorrespondenceNonHermitian2019,%
ozdemirParityTimeSymmetry2019a,%
mirrahimiDynamicallyProtectedCatqubits2014,%
kochControllingOpenQuantum2016,%
zouBellstateGenerationSpin2022,%
arkhipovTopologicalStatePermutations2025a,%
wangAdjustableEntanglementEnhancement2025a,%
kastoryanoDissipativePreparationEntanglement2011}.
A particularly striking phenomenon in this context is chiral state transfer
(CST), where adiabatic encircling of an exceptional point (EP) leads to
direction-dependent state conversion~\cite{kumarNearunitEfficiencyChiral2021c,%
renChiralControlQuantum2022a,%
abbasiTopologicalQuantumState2022b,%
dembowskiObservationChiralState2003a,%
chenDecoherenceInducedExceptionalPoints2022c,%
xuTopologicalEnergyTransfer2016d,%
nasariObservationChiralState2022,%
sunChiralStateTransfer2023a,%
hassanChiralStateConversion2017a,%
uzdinObservabilityAsymmetryAdiabatic2011c,%
zhangTopologicalEigenvalueBraiding2025a,%
gaoPhotonicChiralState2025b,%
wuWuEfficientControlledSymmetric2026}.

Although CST has been extensively studied in classical and semiclassical
settings, a fully quantum realization targeting Bell states has only recently
emerged. In particular, Khandelwal \textit{et
al.}~\cite{khandelwalChiralBellStateTransfer2024} demonstrated that dissipative
Liouvillian dynamics can realize chiral transfer of odd-parity Bell states
$|\Psi^{\pm}\rangle$ within the single-excitation subspace.

Extending CST to the even-parity Bell states $|\Phi^{\pm}\rangle = (|00\rangle
\pm |11\rangle)/\sqrt{2}$ is physically nontrivial. Unlike $|\Psi^{\pm}\rangle$,
which reside entirely in the single-excitation manifold, $|\Phi^{\pm}\rangle$
involve coherence between the zero- and two-excitation sectors. Their generation
and control therefore require a distinct form of engineered dissipation and
coherent coupling. In experimentally relevant platforms such as superconducting
circuits~\cite{leghtasStabilizingBellState2013,%
shankarAutonomouslyStabilizedEntanglement2013}, the coherence between
$|00\rangle$ and $|11\rangle$ is susceptible to decoherence channels distinct
from those affecting odd-parity Bell states. Addressing this regime typically
calls for reservoir engineering based on two-photon processes, such as
parametrically induced nonlinear couplings and engineered two-photon
dissipation, rather than the single-excitation-conserving dynamics used for
$|\Psi^{\pm}\rangle$~\cite{leghtasStabilizingBellState2013,%
shankarAutonomouslyStabilizedEntanglement2013,%
mirrahimiDynamicallyProtectedCatqubits2014,%
ruizTwophotonDrivenKerr2023,%
malekakhlaghQuantumRabiModel2019,%
beaulieuObservationFirstSecondorder2025}.

The present problem therefore differs qualitatively from odd-parity Bell-state
transfer in the single-excitation manifold. The even-parity Bell states
$|\Phi^{\pm}\rangle$ require control of coherence between the zero- and
two-excitation sectors and hence a different dissipative architecture based on
two-photon driving and loss. This change alters both the effective non-Hermitian
structure and the experimental resource requirements and cannot be captured by a
simple change of Bell basis. A central question is whether EP-induced chiral
transfer and dissipative entanglement generation remain possible in this
distinct even-parity setting and, if so, how their dynamical features are
modified.

In this work, we extend the concept of dissipative chiral Bell-state control
from the odd-parity sector to the even-parity manifold, providing a framework
for generating and transferring entanglement. We show that a two-qubit system
with two-photon driving and engineered two-photon loss supports a second-order
exceptional point within the $\{|00\rangle, |11\rangle\}$ subspace, and that
encircling this point induces chiral transfer between $|\Phi^{+}\rangle$ and
$|\Phi^{-}\rangle$. We further show that, when the same loop is initialized in
the separable state $|00\rangle$ at its resonant starting point, the protocol
generates strong even-parity entanglement. This places chiral transfer and
entanglement generation within a common hybrid-Liouvillian description and
clarifies how both phenomena evolve from the conditional no-jump limit toward
the full Lindblad dynamics.

\section{Model}

\begin{figure}[htbp]
	\begin{minipage}[t]{0.56\linewidth}
		\rlap{(a)\hspace{0.5em}}%
		\includegraphics[width=\linewidth]{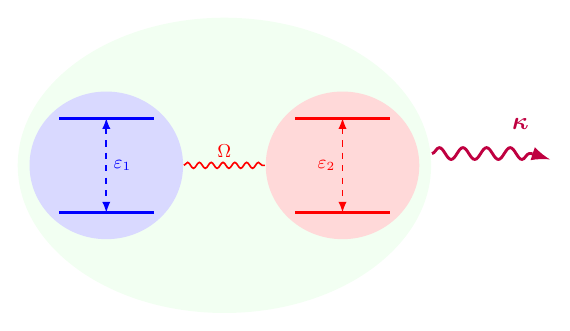}
		\label{subfig:a}
	\end{minipage}
	\hfill
	\begin{minipage}[t]{0.42\linewidth}
		\rlap{(b)\hspace{0.5em}}%
		\includegraphics[width=\linewidth]{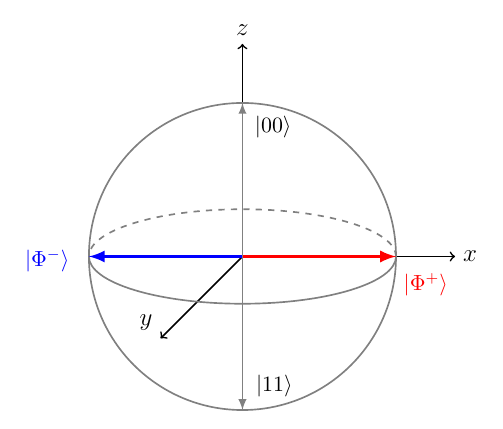}
		\label{subfig:b}
	\end{minipage}
	
	\vspace{0.5em}
	
	\begin{minipage}[t]{0.40\linewidth}
		\rlap{(c)\hspace{0.5em}}%
		\includegraphics[width=\linewidth]{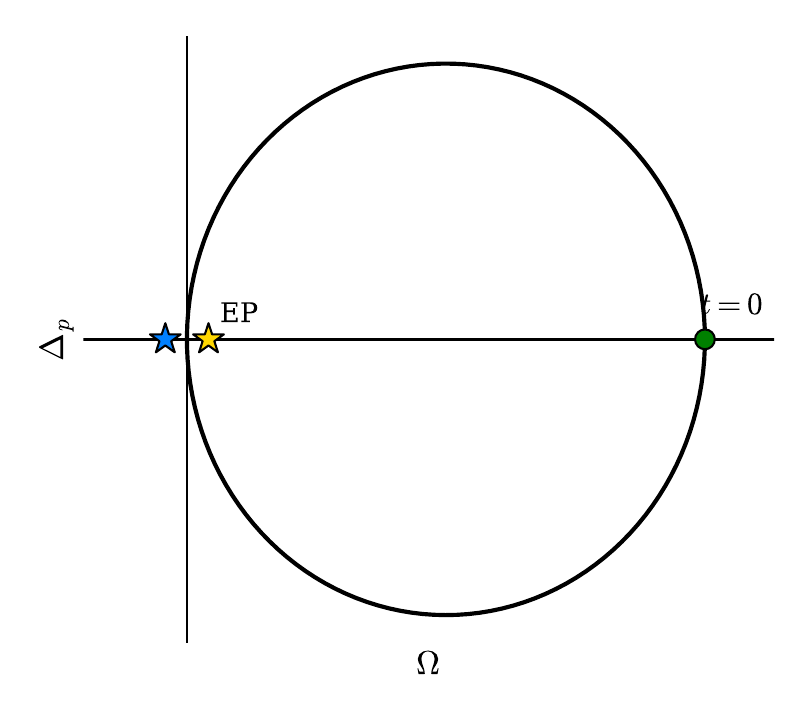}
		\label{subfig:c}
	\end{minipage}
	\hfill
	\begin{minipage}[t]{0.55\linewidth}
		\rlap{(d)\hspace{0.5em}}%
		\includegraphics[width=\linewidth]{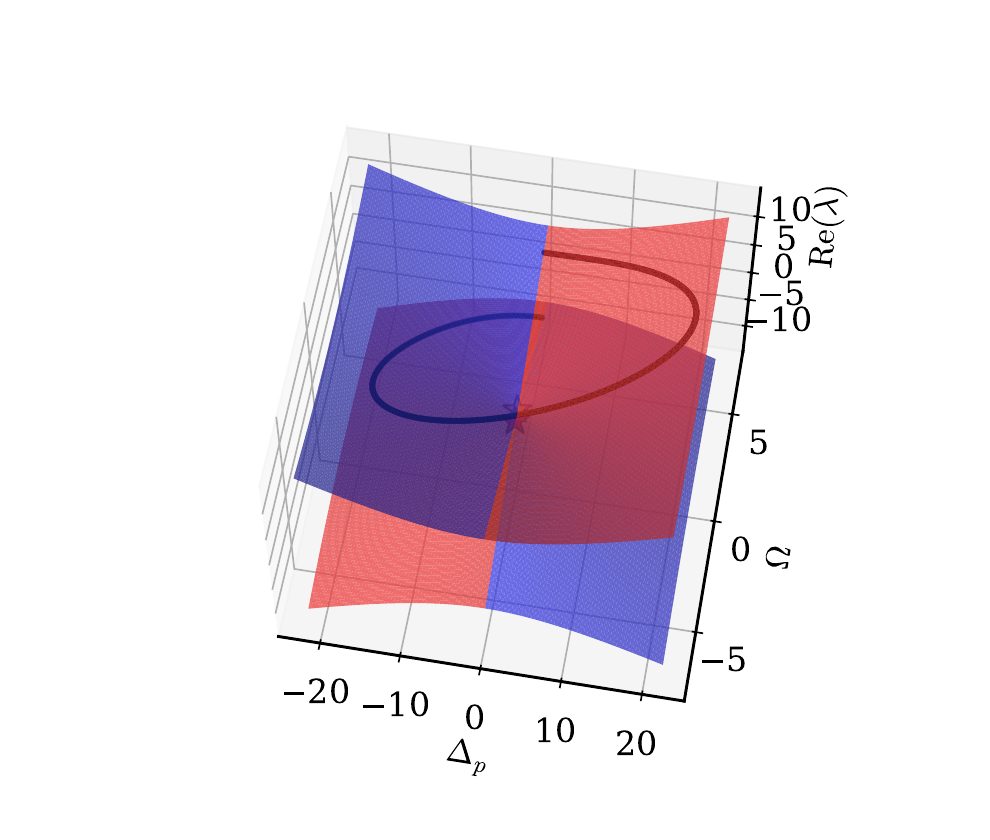}
		\label{subfig:d}
	\end{minipage}
	
	\caption{(a) A system composed of two interacting qubits with two-photon
dissipation into the environment at rate $\kappa$.
		(b) Bloch-sphere representation of the even-parity subspace spanned by
$|11\rangle$ and $|00\rangle$, where $z = |11\rangle\langle11| -
|00\rangle\langle00|$. The target Bell states $|\Phi^\pm\rangle$ are marked on
the sphere.
		(c) Clockwise (CW) and counterclockwise (CCW) evolution trajectories of
$\Omega(t)$ and $\Delta_p(t)$ in the parameter space $(\Omega, \Delta_p)$. The
asterisk marks the second-order exceptional point (EP) of
$H_{\mathrm{even}}^{\mathrm{eff}}$.
		(d) Riemann surfaces corresponding to the eigenvalues of
$H_{\mathrm{even}}^{\mathrm{eff}}$. The branch with stronger decay (more
negative imaginary part) is shown in red, and the black trajectories correspond
to the parameter loop $(\Omega(t),\Delta_p(t))$.}
	\label{fig:diagram}
\end{figure}

Consider a two-photon driven qubit system, as shown in
Fig.~\ref{fig:diagram}(a). In the rotating frame, the system Hamiltonian can be
written as
\begin{equation}
	\begin{aligned}
		H_{\text{even}} =& \sum_{j=1,2} \Delta_j \sigma_+^{(j)} \sigma_-^{(j)} \\
		&+ \Omega \left( \sigma_+^{(1)} \sigma_+^{(2)} e^{-i\phi} + \sigma_-^{(1)}
\sigma_-^{(2)} e^{i\phi} \right),
	\end{aligned}
	\label{eq:hamiltonian_even}
\end{equation}
where $\Delta_j = \varepsilon_j - \omega_d/2$ is the single-qubit detuning in
the rotating frame, $\omega_d$ is the drive frequency, and $\sigma_+^{(j)}$ and
$\sigma_-^{(j)}$ are the raising and lowering operators of the $j$-th qubit,
respectively. The parameter $\Omega$ denotes the two-photon driving strength,
while $\phi$ is the drive phase, which sets the orientation of the transverse
effective field in the even-parity pseudospin space.

We consider a two-photon jump operator
\[
L = \sqrt{\kappa}\,\sigma_{-}^{(1)} \sigma_{-}^{(2)},
\] 
which describes the two-photon loss process from $|11\rangle$ to $|00\rangle$
with rate $\kappa$. According to quantum trajectory theory, during intervals
without quantum jumps, the system evolves under the effective non-Hermitian
Hamiltonian
\begin{equation}
	H_{\text{eff}} = H_{\text{even}} - \frac{i}{2} L^\dagger L.
	\label{eq:effective_hamiltonian}
\end{equation}

Under the Markov approximation, the system dynamics are described by the
Lindblad master equation (setting $\hbar =
1$)~\cite{breuerTheoryOpenQuantum2002b}:
\begin{equation}
	\dot{\rho} = \mathcal{L} \rho = -i\left( H_{\text{eff}} \rho - \rho
H_{\text{eff}}^\dagger \right) + \sum_k L_k \rho L_k^\dagger .
	\label{eq:liouvillian}
\end{equation}
The first term on the right-hand side represents the continuous no-jump
evolution generated by $H_{\text{eff}}$, whereas the second term is the
recycling term associated with quantum
jumps~\cite{harocheExploringQuantumAtoms2006a}. Within the quantum-trajectory
picture~\cite{daleyQuantumTrajectoriesOpen2014a,%
carmichaelQuantumTrajectoryTheory1993a}, this term accounts for the stochastic
state update induced by jump events.

To investigate the evolution under partial post-selection, we introduce a hybrid
Liouvillian equation with an adjustable parameter $q \in [0, 1]$:
\begin{equation}
	\dot{\rho} = \mathcal{L}_{[q]} \rho = -i\left( H_{\text{eff}} \rho - \rho
H_{\text{eff}}^\dagger \right) + q \sum_k L_k \rho L_k^\dagger .
	\label{eq:liouvillian_q}
\end{equation}
Here, $q$ controls the weight of the recycling term. When $q = 0$,
Eq.~\eqref{eq:liouvillian_q} reduces to conditional no-jump dynamics generated
by the non-Hermitian effective Hamiltonian; when $q = 1$, it recovers the
standard Lindblad master equation with the full recycling term. For intermediate
values $0 < q < 1$, the hybrid Liouvillian has a clear physical interpretation
in terms of quantum
trajectories~\cite{mingantiHybridLiouvillianFormalismConnecting2020a}: the
quantum jumps are collected by two perfect detectors with probabilities $q$ and
$1-q$, respectively, and one postselects the trajectories in which the second
detector never clicks. Equivalently, $q$ can be related to a single detector of
finite efficiency $\eta = 1-q$, followed by postselection of the null-detection
outcomes. Intermediate values of $q$ therefore describe \emph{imperfect
monitoring followed by postselection}, rather than partial postselection of
individual jump events or a purely formal interpolation, which also means that
$q$ is not an independent control knob in experiment but is fixed by the
efficiency $\eta=1-q$ of the detection chain that monitors the engineered
two-photon loss. In numerical simulations, we therefore integrate the dynamics
using a time-discretized scheme and renormalize the density matrix at each step
to maintain physical state normalization, which is equivalently described by the
nonlinear master equation
\[
\dot{\rho} = \mathcal{L}_{[q]} \rho + (1-q)\mathrm{Tr}(L\rho L^\dagger)\rho.
\]
With this convention, all fidelities and concurrences reported below for $q<1$
are evaluated on the renormalized conditional state and quantify the quality of
the post-selected state, whereas the case $q=1$ describes unconditional Lindblad
evolution.

By continuously varying $q$, we can track how the dynamics evolves from the
no-jump limit toward the full Lindblad description. This hybrid formalism is
particularly useful for analyzing exceptional-point-induced dynamics in open
systems~\cite{mingantiHybridLiouvillianFormalismConnecting2020a}.

EPs are spectral singularities of non-Hermitian systems. A characteristic
feature of these points is that, when system parameters are varied adiabatically
along a closed path around an EP, different encircling directions can lead to
chiral state or population transfer because of the asymmetry associated with
non-Hermitian gain and loss~\cite{liuDynamicallyEncirclingExceptional2021c,%
abbasiTopologicalQuantumState2022b,%
dopplerDynamicallyEncirclingExceptional2016e}.
The Riemann-sheet structure shown in Fig.~\ref{fig:diagram}(d) illustrates this
mechanism and motivates the directional quantum-state control studied below.

Since both the coherent drive and the dissipation preserve parity, the
even-parity subspace $\mathcal{H}_{\mathrm{even}}=\{|11\rangle,|00\rangle\}$
remains strictly decoupled from the odd-parity subspace
$\{|01\rangle,|10\rangle\}$ throughout the evolution. The effective dynamics can
therefore be restricted to
$\mathcal{H}_{\mathrm{even}}$. It is then convenient to introduce pseudospin
operators acting in this subspace,
\begin{equation}
	\begin{aligned}
		\tau_x &= |11\rangle\langle 00| + |00\rangle\langle 11|, \\
		\tau_y &= -i|11\rangle\langle 00| + i|00\rangle\langle 11|, \\
		\tau_z &= |11\rangle\langle 11| - |00\rangle\langle 00|,
	\end{aligned}
	\label{eq:pseudospin_operators}
\end{equation}
together with the ladder operators
\[
\tau_+ = \frac{\tau_x+i\tau_y}{2} = |11\rangle\langle 00|,\qquad
\tau_- = \frac{\tau_x-i\tau_y}{2} = |00\rangle\langle 11|.
\]
In this pseudospin representation, the effective non-Hermitian Hamiltonian in
the even-parity subspace takes the form
\begin{equation}
	H_{\text{even}}^{\text{eff}} = \left( \frac{\Delta_p}{2} - \frac{i\kappa}{4}
\right) \tau_z + \Omega \left( \cos\phi \, \tau_x + \sin\phi \, \tau_y \right),
	\label{eq:heven_eff_tau}
\end{equation}
where $\Delta_p = (\varepsilon_1 + \varepsilon_2) - \omega_d$ is the two-photon
detuning, which determines the effective field along the $z$-direction of the
pseudospin. The constant term $\frac{\Delta_p}{2} - \frac{i\kappa}{4}$ has been
omitted in Eq.~\eqref{eq:heven_eff_tau}, since it contributes only an overall
energy shift and uniform decay and therefore does not affect the eigenstate
structure or the location of the exceptional point.

Under the resonance condition $\Delta_p = 0$, non-Hermitian dissipation ($\kappa
\neq 0$) induces both a relative phase shift and an amplitude imbalance in the
superposition coefficients, so the instantaneous eigenstates are generally
deformed away from the ideal Bell basis. In the strong-coupling regime $\Omega
\gg \kappa$, however, these eigenstates approach the standard even-parity Bell
states
\[
|\Phi^{\pm}\rangle = \frac{1}{\sqrt{2}}(|00\rangle \pm |11\rangle),
\]
with fidelity $\mathcal{F} > 99\%$. These Bell states are shown on the
pseudospin Bloch sphere in Fig.~\ref{fig:diagram}(b), where $|\Phi^{+}\rangle$
and $|\Phi^{-}\rangle$ lie on the $+x$ and $-x$ axes, respectively, while
$|00\rangle$ and $|11\rangle$ correspond to the north and south poles.

To investigate the dynamical consequences of the EP structure, we evolve the
system along a closed path in parameter space that encloses the singularity.
Since the EP is determined by the two real control parameters $\Omega$ and
$\Delta_p$, where tuning $\Delta_p$ corresponds to sweeping the two-photon drive
frequency across resonance and tuning $\Omega$ corresponds to modulating the
drive amplitude, we choose the periodic parameter loop
\begin{equation}
	\begin{split}
		\Delta_p(t) &= \Delta_0 \pm R_\Delta \sin\left( \frac{2\pi t}{T} \right), \\
		\Omega(t) &= \Omega_0 + R_\Omega \cos\left( \frac{2\pi t}{T} \right),
	\end{split}
	\label{eq:parameter_loop}
\end{equation}
where $(\Delta_0,\Omega_0)$ is the center of the parameter trajectory, and $T$
is the driving period.
The ``$+$'' and ``$-$'' signs in $\Delta_p(t)$ correspond to clockwise (CW) and
counterclockwise (CCW) encircling, respectively. For the effective Hamiltonian
in Eq.~\eqref{eq:heven_eff_tau}, the second-order exceptional point is located
at $\Omega = \pm\kappa/4$ and $\Delta_p = 0$. By choosing the modulation
amplitudes $R_\Delta$ and $R_\Omega$ appropriately, the trajectory can be made
to encircle this exceptional point, thereby enabling chiral dynamics governed by
the combined effect of branch exchange and non-Hermitian gain/loss asymmetry.
For the working point used below ($\Omega_0 = 3\kappa$, $R_\Omega = 3\kappa$,
$R_\Delta = 20\kappa$), the loop encloses exactly one of the two EPs, namely the
one at $\Omega = +\kappa/4$, so that the argument of the square root in the
eigenvalues winds once around the origin along the loop (winding number $\mp 1$
for CW/CCW) and the two eigenvalue branches exchange after one cycle. Note that
this requires $-\kappa/4 < \Omega_0 - R_\Omega < \kappa/4$; a loop that enclosed
both exceptional points ($\Omega_0 - R_\Omega < -\kappa/4$) or neither of them
($\Omega_0 - R_\Omega > \kappa/4$) would have zero net winding, and the branch
exchange would be suppressed.

\section{Chiral Bell State Transfer}

Let the initial state of the system be $\rho(0)$. The time-evolved density
matrix at time $t$ can be formally written as
\begin{equation}
	\rho(t) = \mathcal{T} \exp\left[ \int_0^t \mathcal{L}_{[q]}(t')\, dt' \right]
\rho(0),
	\label{eq:rho_t}
\end{equation}
where $\mathcal{T}$ denotes the time-ordering operator and $\mathcal{L}_{[q]}$
is the Liouvillian superoperator introduced in Sec.~II. For $0 \le q < 1$, the
linear evolution generated by $\mathcal{L}_{[q]}$ is not trace preserving, so in
numerical calculations we integrate the dynamics using a time-discretized scheme
and renormalize the density matrix at each step to maintain physical state
normalization. Accordingly, the Bell-state fidelity is evaluated for the
normalized state as
\begin{equation}
	\mathcal{F}_{|\Phi^\pm\rangle}(t) :=
\mathrm{Tr}\left[|\Phi^\pm\rangle\langle\Phi^\pm|\,\rho(t)\right],
\end{equation}
which quantifies the overlap of the evolving conditional state with the target
even-parity Bell states.

\begin{figure}[htbp]
	\begin{minipage}[t]{0.45\linewidth}
		\rlap{(a)\hspace{0.5em}}%
		\includegraphics[width=\linewidth]{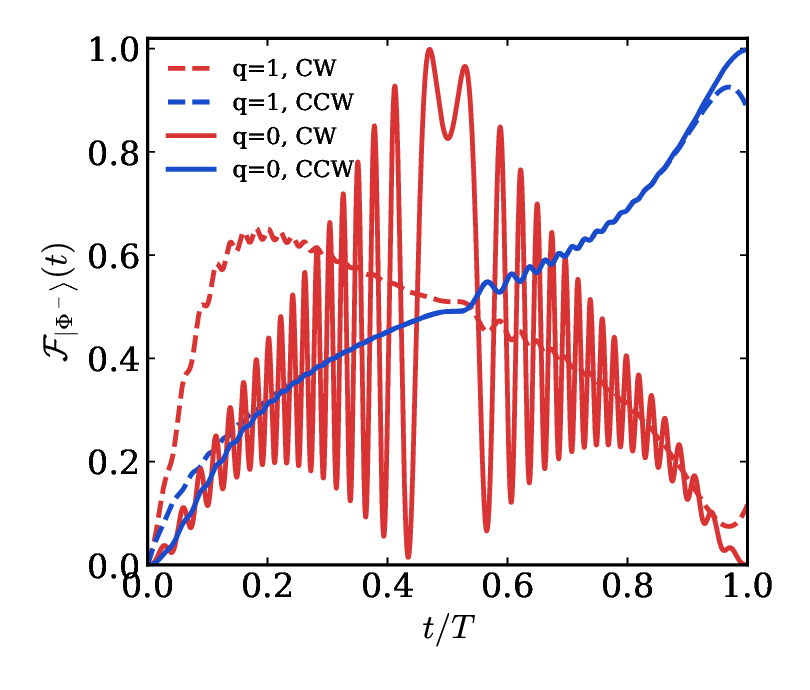}
		\label{subfig:a}
	\end{minipage}
	\hfill
	\begin{minipage}[t]{0.45\linewidth}
		\rlap{(b)\hspace{0.5em}}%
		\includegraphics[width=\linewidth]{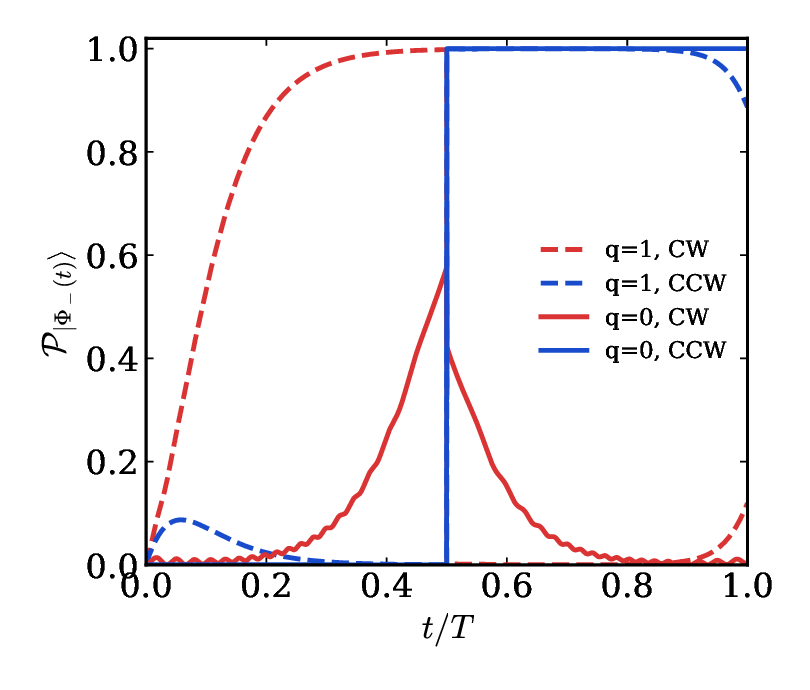}
		\label{subfig:b}
	\end{minipage}
	
		\vspace{0.5em}
		
	\begin{minipage}[t]{0.45\linewidth}
		\rlap{(c)\hspace{0.5em}}%
		\includegraphics[width=\linewidth]{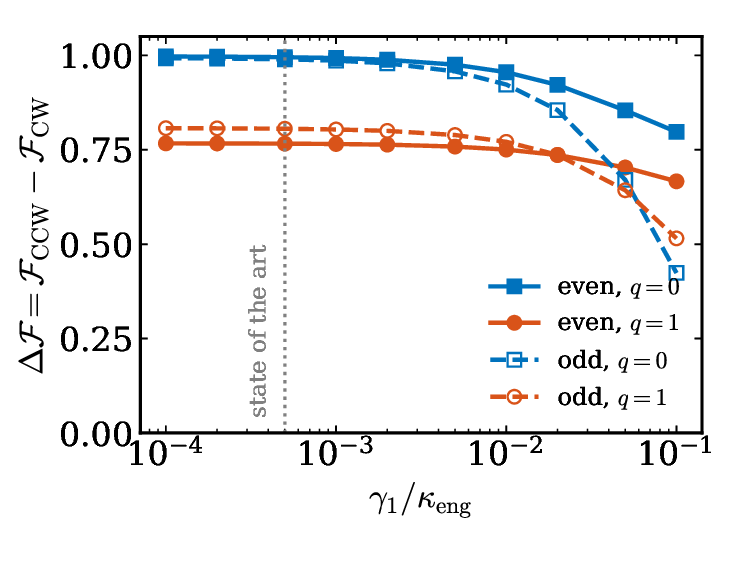}
		\label{subfig:c}
	\end{minipage}
	\hfill
	\begin{minipage}[t]{0.45\linewidth}
		\rlap{(d)\hspace{0.5em}}%
		\includegraphics[width=\linewidth]{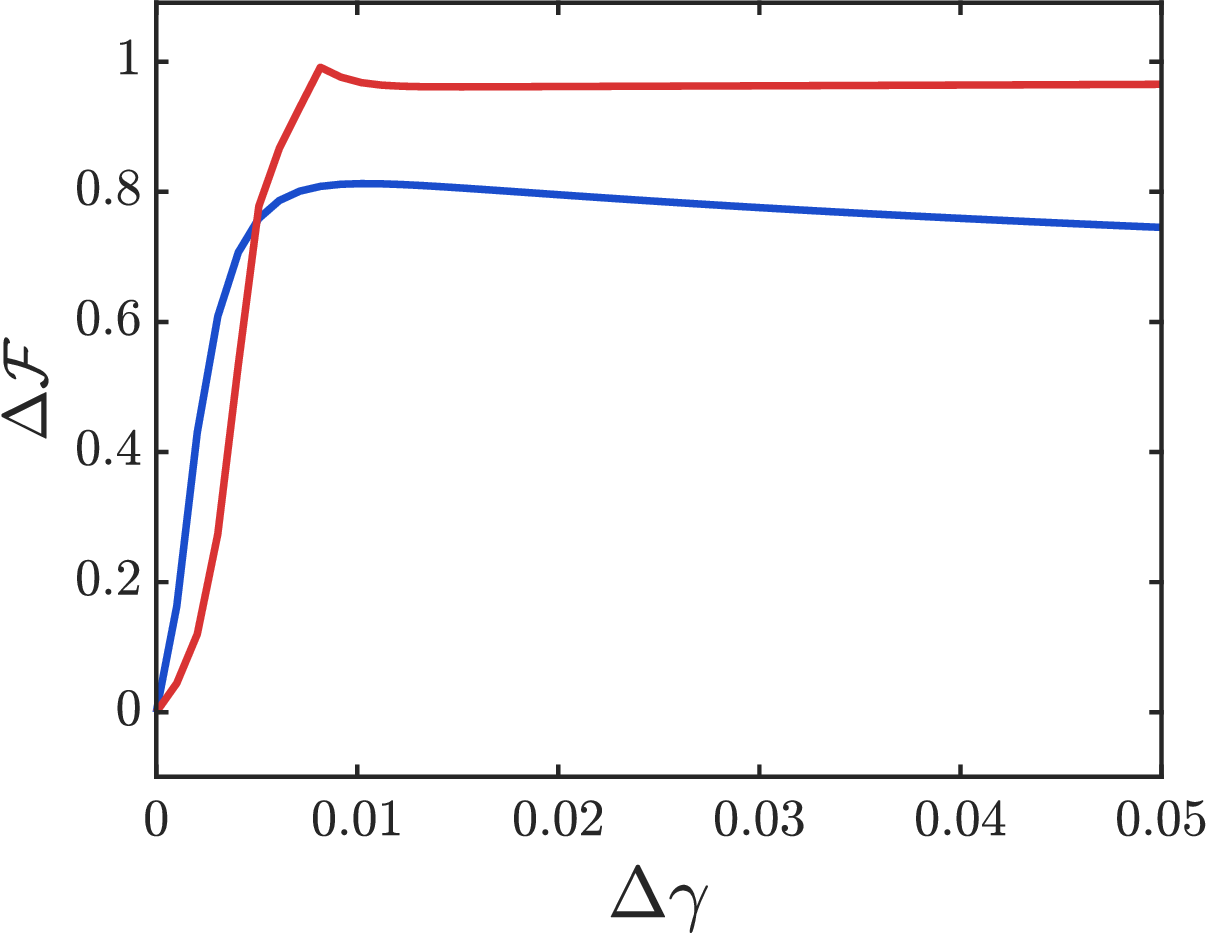}
		\label{subfig:d}
	\end{minipage}
	\caption{(a) Time evolution of the fidelity $\mathcal{F}_{|\Phi^-\rangle}(t)$
for a protocol that starts from $|\Phi^+\rangle$ and targets $|\Phi^-\rangle$.
The red and blue curves correspond to CW and CCW
encircling, respectively. 
		(b) Projection probability of the evolving state onto the instantaneous
eigenstate $|\Phi_-(t)\rangle$. 
		Panels (c) and (d) denote robustness of the chiral effect \(\Delta \mathcal{F}
=
\mathcal{F}_{\text{CCW}} - \mathcal{F}_{\text{CW}}\).
		(c) \(\Delta\mathcal{F}\) when the present even-parity scheme (solid lines,
filled markers) and the odd-parity scheme of
Ref.~\cite{khandelwalChiralBellStateTransfer2024} (dashed lines, open markers)
are subjected to the same parasitic perturbation: an unmonitored local
single-photon loss of rate \(\gamma_1\) on each qubit, expressed in units of the
engineered dissipation scale \(\kappa_{\mathrm{eng}}\) of each scheme
(\(\kappa_{\mathrm{eng}}=\kappa\) for the even-parity protocol and
\(\kappa_{\mathrm{eng}}=\Delta\gamma\) for the odd-parity one). Squares: full
postselection (\(q=0\)); circles: unconditional dynamics (\(q=1\)). The vertical
dotted line marks the state-of-the-art ratio \(\gamma_1/\kappa \approx
5\times10^{-4}\) for superconducting
circuits~\cite{placeNewMaterialPlatform2021}.
		(d) \(\Delta\mathcal{F}\) of the odd-parity scheme as a function of its
dissipation modulation amplitude \(\Delta\gamma\) (in units of the qubit
transition energy \(\varepsilon\)), for \(q=0\) and \(q=1\).
				The odd-parity simulations follow the model of
Ref.~\cite{khandelwalChiralBellStateTransfer2024} with
\(\Delta\delta/\varepsilon = 0.04\), \(g/\varepsilon = 0.01\), \(\alpha = 1.2\),
and \(T\varepsilon = 2500\); panel (c) uses \(\Delta\gamma/\varepsilon =
0.008\). 
				Parameters of the even-parity scheme: $\kappa = 1$, $\phi = 0$, $\Omega_0 =
3\kappa$, $R_\Omega = 3\kappa$, $\Delta_0 = 0$, $R_\Delta = 20\kappa$, and
$T\kappa = 15$.}
	\label{fig:F_P_t}
\end{figure}

Figure~\ref{fig:F_P_t}(a) shows the time evolution of the fidelity
$\mathcal{F}_{|\Phi^-\rangle}(t)$ for a protocol that starts from the Bell state
$|\Phi^+\rangle$ and targets $|\Phi^-\rangle$. Under full post-selection
($q=0$), the system exhibits pronounced chiral transfer at $t=T$: along the CCW
path, the initial state $|\Phi^+\rangle$ is converted into the target state
$|\Phi^-\rangle$ with a fidelity of $0.9976$, whereas along the CW path the
final fidelity is only $0.0002$. Under the full Lindblad evolution ($q=1$), the
chiral contrast is reduced but remains clearly visible.

The physical origin of this chiral response is reflected in the projection
probability $\mathcal{P}_{|\Phi_-(t)\rangle}(t)$ of the evolving state onto the
instantaneous eigenstate $|\Phi_-(t)\rangle$ of the effective Hamiltonian
$H_{\text{eff}}$, shown in Fig.~\ref{fig:F_P_t}(b). For CCW evolution with
$q=0$, this projection probability increases from a near-zero value at $t=0$ to
$1.000$ at $t=T$, indicating that the dynamics becomes locked onto the target
eigenstate. Under CW evolution, although the projection probability fluctuates
during the cycle, it remains close to zero at the end of the loop. This marked
directional asymmetry shows that the CCW path efficiently drives the system onto
the $|\Phi_-(t)\rangle$ branch, whereas the CW path suppresses this transfer,
yielding a clear chiral response governed by the combined effect of branch
exchange and non-Hermitian gain/loss asymmetry.

Note, however, that the chiral response does not hinge on the topological
enclosure itself. If the modulation amplitude is reduced to $R_\Omega =
2\kappa$, so that the loop encloses neither EP and the eigenvalue branches no
longer exchange, the fully postselected chiral contrast remains essentially
unchanged ($\Delta\mathcal{F} = 0.997$ for both working points, and $0.79$
versus $0.77$ at $q=1$), which is consistent with chiral state transfer without
encircling an EP~\cite{nasariObservationChiralState2022,%
sunChiralStateTransfer2023a,%
hassanChiralStateConversion2017a} and with the parameter robustness reported for
the odd-parity scheme~\cite{khandelwalChiralBellStateTransfer2024}. The
directional asymmetry is thus dominated by the time-integrated low-loss branch
selection analyzed in Appendix~A, whereas enclosing the EP only decides whether
a genuine branch exchange accompanies the transfer.

Figures~\ref{fig:F_P_t}(c) and~\ref{fig:F_P_t}(d) illustrate the robustness of
the chiral effect \(\Delta \mathcal{F} = \mathcal{F}_{\text{CCW}} -
\mathcal{F}_{\text{CW}}\) against a parasitic decoherence channel and against
the dissipation modulation amplitude, respectively.

In Fig.~\ref{fig:F_P_t}(c), we compare the present even-parity scheme with the
odd-parity scheme of Ref.~\cite{khandelwalChiralBellStateTransfer2024} under the
same perturbation, namely an unmonitored parasitic single-photon loss of rate
\(\gamma_1\) acting locally on each qubit, where \(\gamma_1\) is measured in
units of the engineered dissipation scale \(\kappa_{\mathrm{eng}}\) of each
scheme (\(\kappa\) for the even-parity protocol and \(\Delta\gamma\) for the
odd-parity one; the corresponding loop durations are comparable, \(T\kappa =
15\) and \(T\Delta\gamma = 20\)). We can see that both schemes display a wide
plateau. The chiral contrast is essentially unaffected for \(\gamma_1 \lesssim
10^{-3}\,\kappa_{\mathrm{eng}}\) and degrades appreciably only when \(\gamma_1\)
reaches the percent level of the engineered rate. At the state-of-the-art ratio
\(\gamma_1/\kappa \approx 5\times10^{-4}\) available in superconducting circuits
(see Sec.~\ref{sec:feasibility}), the reduction of \(\Delta\mathcal{F}\) is
below \(0.3\%\) for both \(q=0\) and \(q=1\). In the strongly perturbed regime,
however, the two schemes respond differently. At \(\gamma_1 =
0.1\,\kappa_{\mathrm{eng}}\) the fully postselected even-parity contrast remains
\(\Delta\mathcal{F} = 0.80\), whereas its odd-parity counterpart drops to
\(0.42\). The physical reason lies in the structure of the two manifolds: in the
even-parity scheme the parasitic channel ultimately returns population to
\(|00\rangle\), which lies \emph{inside} the protected manifold, whereas in the
odd-parity scheme it removes population to \(|00\rangle\), \emph{outside} the
single-excitation manifold, from which the postselected dynamics cannot recover
it.

As shown in Fig.~\ref{fig:F_P_t}(d), the dependence of \(\Delta \mathcal{F}\) on
\(\Delta\gamma\) in the odd-parity scheme exhibits two distinct stages. For
\(\Delta\gamma \lesssim 0.01\,\varepsilon\), \(\Delta \mathcal{F}\) increases
steeply with \(\Delta\gamma\), because as \(\Delta\gamma \to 0\) the parameter
trajectory degenerates into a one-dimensional line segment that no longer
encloses the EP, and the chirality disappears. Once \(\Delta\gamma\) exceeds
this threshold, the loop encloses the EP and the chiral contrast saturates
rapidly, remaining robust against further changes of \(\Delta\gamma\). This
behavior is consistent with the established picture that the chiral response is
controlled by the interplay between the geometry of the closed trajectory and
the Riemann-sheet structure it samples~\cite{nasariObservationChiralState2022}.

Table~\ref{tab:comparison} summarizes the comparison between the two schemes in
terms of mechanism, required resources, and robustness. In both cases the
transfer is governed by branch selection near a second-order EP of the effective
non-Hermitian Hamiltonian; the schemes differ in the manifold hosting the EP, in
the coherent and dissipative resources that engineer it, and in how quantum
jumps act on that manifold.

\begin{table}[htbp]
	\caption{Comparison between the even-parity scheme studied in this work and the
odd-parity scheme of Ref.~\cite{khandelwalChiralBellStateTransfer2024}. The last
two rows quote the chiral contrast $\Delta\mathcal{F}$ obtained with the loop
parameters of Figs.~\ref{fig:F_P_t}(c) and (d).}
	\label{tab:comparison}
	\begin{ruledtabular}
		\begin{tabular}{p{0.30\linewidth}p{0.30\linewidth}p{0.28\linewidth}}
			& Even parity (this work) & Odd parity
(Ref.~\cite{khandelwalChiralBellStateTransfer2024}) \\
			\colrule
			Bell states & $|\Phi^\pm\rangle$ (zero- and two-excitation coherence) &
$|\Psi^\pm\rangle$ (single-excitation) \\
			Working manifold & $\{|00\rangle,|11\rangle\}$, exactly decoupled by parity &
$\{|01\rangle,|10\rangle\}$, not closed under jumps \\
			Coherent resource & parametric two-photon drive $\Omega$ &
excitation-exchange coupling $g$ \\
			Dissipative resource & engineered two-photon loss $\kappa$ & engineered local
gain and loss $\gamma_{1,2}(t)$ \\
			Engineered jumps & act within the manifold ($L \propto \tau_-$) & connect the
manifold to the $|00\rangle$, $|11\rangle$ sectors \\
			Parasitic loss $\gamma_1$ & recycles population into the manifold
($\to|00\rangle$) & removes population from the manifold \\
			$\Delta\mathcal{F}$, ideal ($q=0$ / $q=1$) & $0.997$ / $0.77$ & $0.99$ /
$0.81$ \\
			$\Delta\mathcal{F}$ at $\gamma_1=0.1\,\kappa_{\mathrm{eng}}$ ($q=0$) & $0.80$
& $0.42$ \\
		\end{tabular}
	\end{ruledtabular}
\end{table}

\begin{figure}[htbp]
	\begin{minipage}[t]{0.45\linewidth}
		\rlap{(a)\hspace{0.5em}}%
		\includegraphics[width=\linewidth]{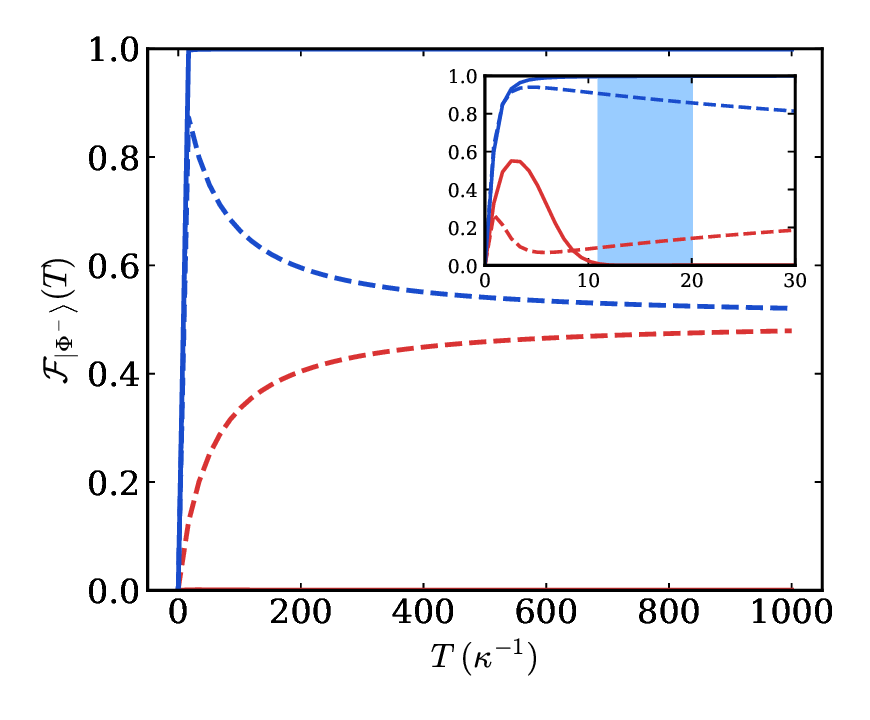}
		\label{subfig:a}
	\end{minipage}
	\hfill
	\begin{minipage}[t]{0.45\linewidth}
		\rlap{(b)\hspace{0.5em}}%
		\includegraphics[width=\linewidth]{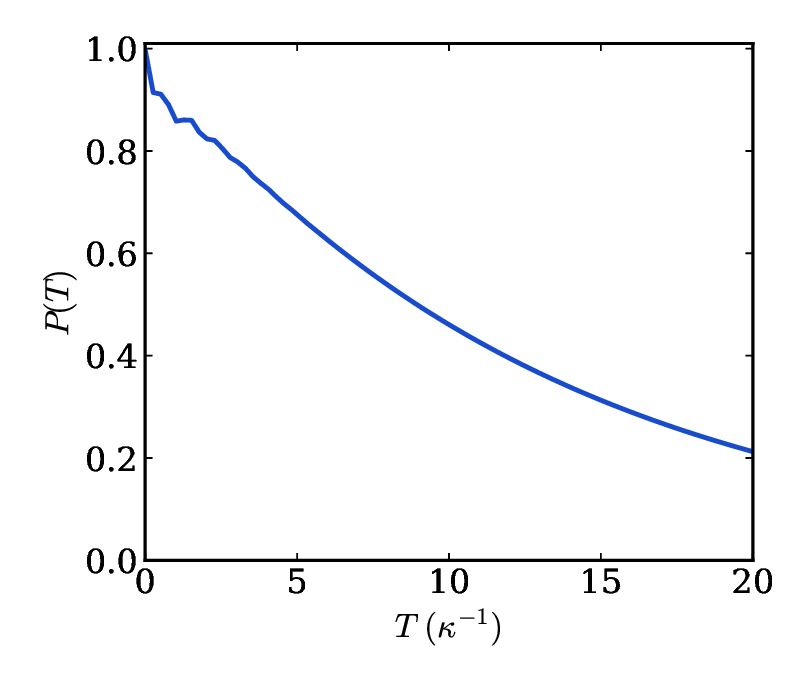}
		\label{subfig:b}
	\end{minipage}
	
	\caption{(a) Final-state fidelity $\mathcal{F}_{|\Phi^-\rangle}(T)$ as a
function of the driving period $T$. The inset shows a magnified view of the
range $T \in [0,30]$, and the light-blue shaded region highlights the
intermediate window $T \in [11,20]\kappa^{-1}$.
		(b) Post-selection success probability $P(T)=\mathrm{Tr}[\rho(T)]$ as a
function of the period $T$ for $q=0$. The other parameters are the same as in
Fig.~\ref{fig:F_P_t}.}
	\label{fig:F_P_T}
\end{figure}

The dependence of this chiral selectivity on the driving period is summarized in
Fig.~\ref{fig:F_P_T}(a), which plots the final-state fidelity
$\mathcal{F}_{|\Phi^-\rangle}(T)$ as a function of $T$. In the long-period
regime ($T \gg \kappa^{-1}$), the fully postselected dynamics ($q=0$) exhibits
strong chiral selectivity: along the CCW path, the fidelity approaches unity,
indicating nearly complete transfer to the target Bell state, whereas along the
CW path it approaches zero, showing that the transfer is effectively suppressed.
By contrast, for $q=1$, the chiral selectivity in the long-period regime is
weakened. The fidelity along the CCW path decreases, while that along the CW
path increases. In both directions, the system tends toward a mixed state, and
no high-contrast chiral transfer survives in the adiabatic limit.

In the short-period regime, the fidelity shows pronounced oscillatory behavior
caused by nonadiabatic transitions. By contrast, the intermediate window $T \in
[11,20]\kappa^{-1}$ retains good chiral selectivity. This regime therefore
provides a favorable operating window in which non-Hermitian adiabaticity and
mode-selection effects are well balanced. Even for $q=1$, the fidelity along the
CCW path remains above $0.857$ throughout the window, while that along the CW
path stays below $0.143$.

While the fully postselected evolution yields near-perfect chiral transfer, it
is also necessary to assess the associated probabilistic cost. As shown in
Fig.~\ref{fig:F_P_T}(b), the post-selection success probability $P(T) =
\mathrm{Tr}\left[\rho(T)\right]$ along the CCW path decreases monotonically with
$T$ for $q=0$. Here $\rho(T)$ denotes the unnormalized conditional state
generated by the linear evolution in Eq.~\eqref{eq:rho_t}, whose trace records
the accumulated no-jump probability; the renormalized state used for all
fidelities is $\rho(T)/P(T)$. At $T=20\kappa^{-1}$, the success probability is
$0.213$; at $T=11\kappa^{-1}$, where the fidelity still displays strong chiral
selectivity, it reaches $0.426$. This trend shows that, within the fully
postselected ($q=0$) dynamics, high-fidelity chiral transfer comes at a
substantial probabilistic cost and makes explicit the trade-off between transfer
fidelity and post-selection success probability in this open-system protocol. In
the unconditional regime ($q=1$), by contrast, the protocol becomes
deterministic at the price of a reduced fidelity. Every experimental run is
retained, so that no probabilistic overhead is incurred. For intermediate values
of $q$, the success probability interpolates monotonically between these two
limits, since only a fraction $1-q$ of the jump flux is discarded by the
postselection. Therefore, the conditional regime ($q=0$) is advantageous when
high-fidelity states are required and probabilistic post-selection is
acceptable, whereas the unconditional regime ($q=1$) is preferable for
deterministic operation in which a moderately reduced fidelity (here $0.857$ at
the edge of the operating window) is sufficient.

\section{Entanglement Generation}

We now turn to entanglement generation from a separable initial state. The
protocol uses the same loop as in Eq.~\eqref{eq:parameter_loop}, initialized in
the separable state $|00\rangle$ at the starting point of the trajectory,
$\Omega(0) = \Omega_0 + R_\Omega$ and $\Delta_p(0) = 0$. At this resonant point
the instantaneous eigenstates of $H_{\text{even}}^{\text{eff}}$ are close to the
Bell states $|\Phi^\pm\rangle$, so that the separable initial state enters the
loop as an almost equal-weight superposition of the two eigenstate branches,
$|00\rangle = (|\Phi^+\rangle + |\Phi^-\rangle)/\sqrt{2}$. Encircling the EP
then drives this superposition onto the low-loss branch through the same
time-integrated branch-selection mechanism that underlies the chiral transfer,
which converts the chiral loop into an entanglement-generation protocol. Because
the dynamics remains strictly confined to the even-parity subspace
$\{|00\rangle, |11\rangle\}$, the density matrix retains a reduced X-type
structure, and the concurrence can be written in the simplified
form~\cite{woottersEntanglementFormationArbitrary1998b}
\begin{equation}
	\mathcal{C}(\rho) := 2 \max\left\{0, |d| - \sqrt{p_{01} p_{10}}\right\},
	\label{eq:concurrence}
\end{equation}
where $p_{01} = \langle 01|\rho|01\rangle$, $p_{10} = \langle
10|\rho|10\rangle$, and the off-diagonal element $d = \langle 00|\rho|11\rangle$
characterizes the coherence between the $|00\rangle$ and $|11\rangle$ levels. In
the ideal protocol, exact parity conservation enforces $p_{01} = p_{10} = 0$, so
that Eq.~\eqref{eq:concurrence} reduces to $\mathcal{C} = 2|d|$; we nevertheless
retain the general X-state form, which remains valid when parity-breaking
parasitic channels are included, as in the robustness analysis of
Fig.~\ref{fig:F_P_t}(c). With this convention, $\mathcal{C}=0$ corresponds to a
separable state, whereas $\mathcal{C}=1$ corresponds to a maximally entangled
state.

\begin{figure}[htbp]
	\centering
	\begin{minipage}[t]{0.45\linewidth}
		\rlap{(a)\hspace{0.5em}}%
		\includegraphics[width=\linewidth]{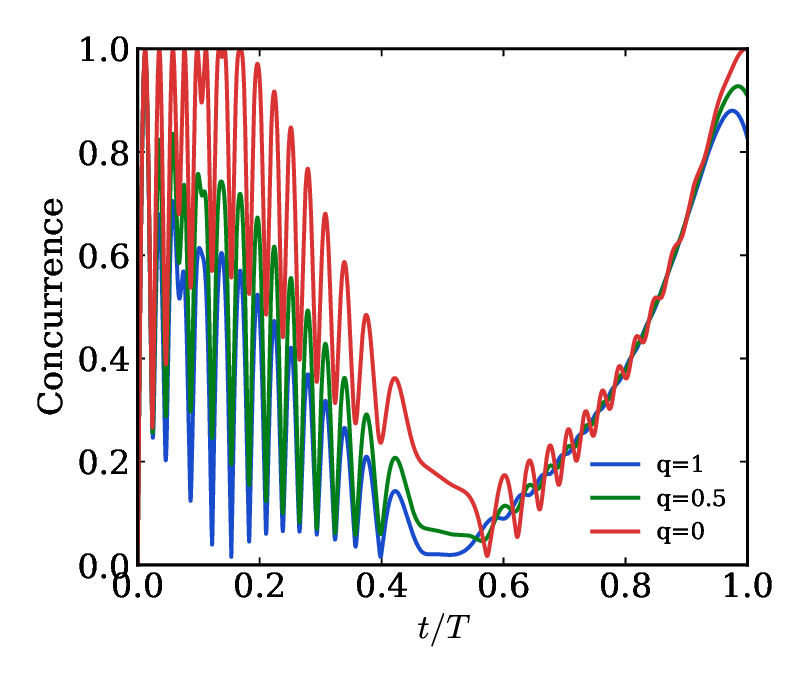}
		\label{subfig:a}
	\end{minipage}
	\hfill
	\begin{minipage}[t]{0.45\linewidth}
		\rlap{(b)\hspace{0.5em}}%
		\includegraphics[width=\linewidth]{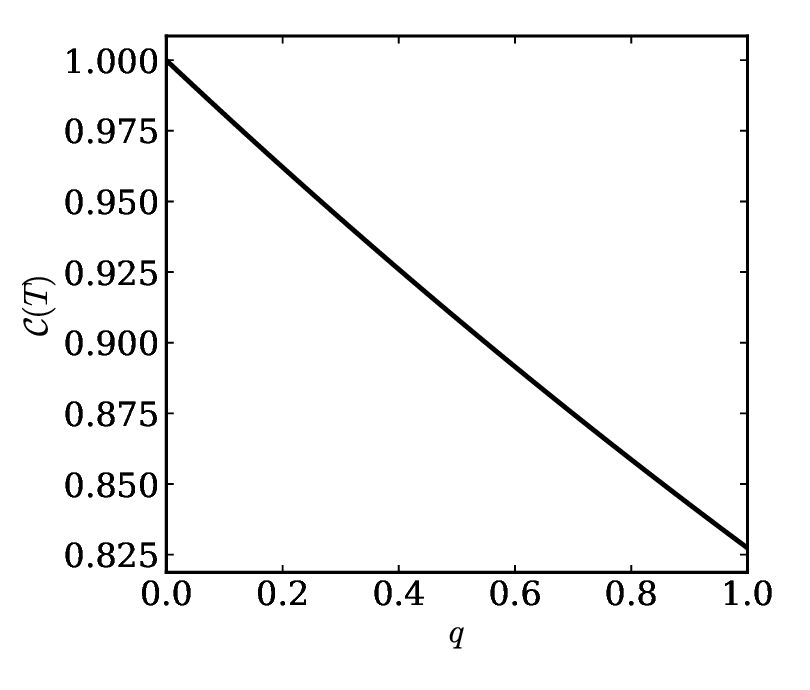}
		\label{subfig:b}
	\end{minipage}
	
	\vspace{0.5em}
	
	\begin{minipage}[t]{0.45\linewidth}
		\rlap{(c)\hspace{0.5em}}%
		\includegraphics[width=\linewidth]{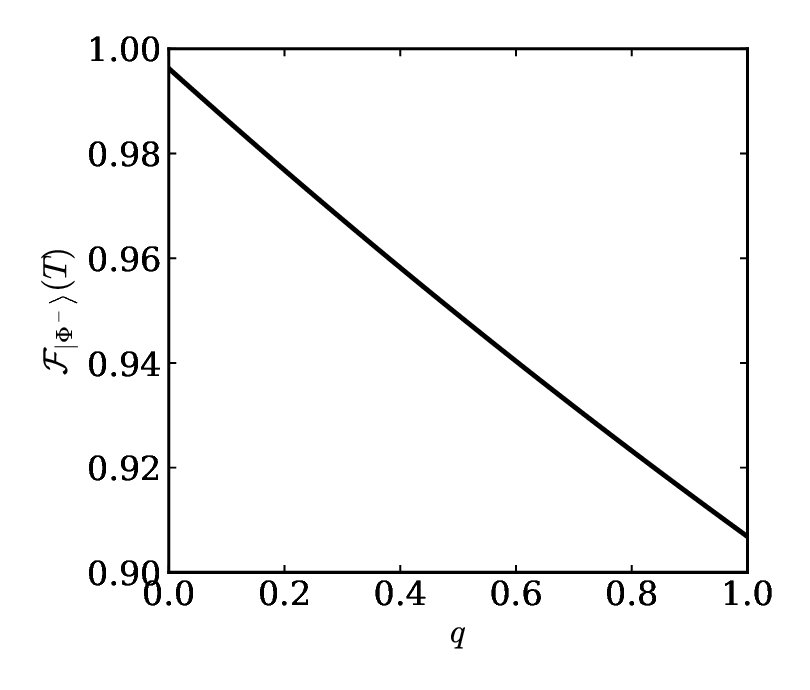}
		\label{subfig:c}
	\end{minipage}
	\hfill
          \begin{minipage}[t]{0.45\linewidth}
		\rlap{(d)\hspace{0.5em}}%
		\includegraphics[width=\linewidth]{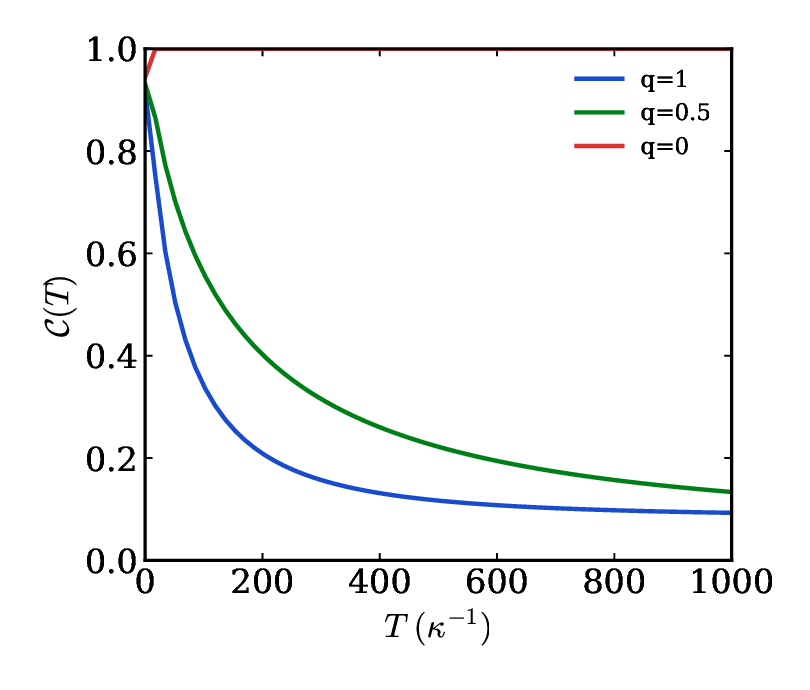}
		\label{subfig:d}
	\end{minipage}
	
	\caption{(a) Time evolution of the concurrence $\mathcal{C}$ as a function of
$t/T$ for CCW loops with $q=0$, $0.5$, and $1$.
		(b) Final concurrence $\mathcal{C}(T)$ as a function of the recycling
parameter $q$ for the CCW loop.
		(c) Final fidelity $\mathcal{F}_{|\Phi^-\rangle}(T)$ as a function of $q$ for
the CCW loop.
		(d) Final concurrence $\mathcal{C}(T)$ as a function of the driving period $T$
for the CCW loop.
		The initial state is $|00\rangle$ in all cases. For panels (a)--(c),
$T\kappa=11$. Other parameters are the same as in Fig.~\ref{fig:F_P_t}.}
	\label{C_F_T_q}
\end{figure}

Figure~\ref{C_F_T_q}(a) shows the time evolution of the concurrence during the
entanglement-generation protocol. Starting from the separable state
$|00\rangle$, the concurrence increases substantially over the course of the
loop and reaches a large final value at $t/T=1$. The generated entanglement is
strongest for $q=0$ and decreases as the recycling contribution increases,
indicating that quantum jumps reduce the attainable entanglement but do not
eliminate it over the timescales considered here.

This trend is quantified in Fig.~\ref{C_F_T_q}(b), which plots the final
concurrence as a function of the recycling parameter $q$. The concurrence
decreases monotonically as the weight of the jump term increases. Even under the
full Lindblad evolution, however, the concurrence still reaches $0.827$, showing
that the protocol retains substantial entanglement-generation capability in the
presence of dissipation. A similar trend is seen in the final-state fidelity
shown in Fig.~\ref{C_F_T_q}(c), which decreases monotonically with $q$ and
indicates a growing deviation from the ideal target state $|\Phi^-\rangle$ as
recycling becomes stronger.

The role of the driving period is shown in Fig.~\ref{C_F_T_q}(d). For
dissipative trajectories ($q=0.5$ and $q=1$), the concurrence decreases
monotonically as $T$ increases. In the long-period regime, the state gradually
approaches a mixed state under the cumulative effect of dissipation, and the
concurrence is correspondingly reduced. This behavior shows that entanglement
generation is intrinsically a finite-time effect in the dissipative setting and
that the driving period must be chosen within an appropriate window to avoid
excessive degradation of the generated entanglement.

Taken together, these results show that the same EP-encircling loop can be used
not only for chiral Bell-state transfer, but also for finite-time generation of
strong even-parity entanglement from a separable initial state. They also
identify a practical operating window in which entanglement generation remains
effective despite dissipative mixing.

\section{Experimental Feasibility}
\label{sec:feasibility}

The two key resources required by our scheme, coherent two-photon driving and
engineered two-photon loss, have both been demonstrated in
superconducting-circuit platforms, in the context of Bell-state
stabilization~\cite{shankarAutonomouslyStabilizedEntanglement2013} and
engineered two-photon dissipation~\cite{leghtasConfiningStateLight2015,%
lescanneExponentialSuppressionBitflips2020}.
All parameters in this work are normalized to the engineered two-photon
dissipation rate $\kappa$. Taking $\kappa/2\pi = 1\,\text{MHz}$ as a
representative value ($\kappa^{-1} \approx 0.16\,\mu\text{s}$), the working
regime $\Omega_0 = 3\kappa$, $R_\Omega = 3\kappa$, $R_\Delta = 20\kappa$, and $T
= 15/\kappa$ translates into driving and modulation amplitudes of a few to a few
tens of MHz and a loop period of $T \approx 2.4\,\mu\text{s}$, which are
achievable with standard parametric driving and flux-biasing techniques. The
two-photon loss can be engineered by dispersively coupling the two qubits to a
common low-$Q$ resonator mode, where a parametric pump converts two qubit
excitations into a single rapidly decaying resonator
photon~\cite{shankarAutonomouslyStabilizedEntanglement2013,%
lescanneExponentialSuppressionBitflips2020}.

The even-parity subspace $\{|00\rangle, |11\rangle\}$ is strictly preserved by
the engineered two-photon drive and two-photon loss. Parasitic decoherence
channels such as single-photon loss and pure dephasing can couple the even- and
odd-parity sectors, but state-of-the-art superconducting transmon qubits achieve
single-photon relaxation times exceeding
$300\,\mu\text{s}$~\cite{placeNewMaterialPlatform2021}, which corresponds to
parasitic loss rates as low as $\gamma_1/2\pi \sim 0.5\,\text{kHz}$, i.e.,
$\gamma_1/\kappa \approx 5\times10^{-4}$, three orders of magnitude below the
engineered dissipation rate. The simulations of Fig.~\ref{fig:F_P_t}(c) confirm
this estimate quantitatively: including a local single-photon loss of rate
$\gamma_1 = 10^{-3}\kappa$ on both qubits reduces the chiral contrast
$\Delta\mathcal{F}$ by only $0.45\%$ ($0.23\%$) for $q=0$ ($q=1$) and the
generated concurrence $\mathcal{C}(T)$ by $1.4\%$ ($0.6\%$), while pure
dephasing of the same strength changes $\Delta\mathcal{F}$ by about $0.3\%$.
Parasitic decoherence at realistic levels therefore has a negligible effect on
the chiral dynamics discussed in this work.

\section{Conclusions}

In conclusion, we have shown that hybrid Liouvillian dynamics can be used to
realize chiral transfer and finite-time dissipative generation of even-parity
Bell states in a two-qubit system with engineered two-photon loss. Encircling
the EP in the $(\Omega, \Delta_p)$ plane induces directional conversion between
$|\Phi^+\rangle$ and $|\Phi^-\rangle$, while the same loop initialized in the
separable state $|00\rangle$ generates strong even-parity entanglement. The
underlying mechanism is dominated by time-integrated low-loss branch selection,
with the branch exchange of exceptional-point encircling fixing the geometric
structure of the evolution.

Within the hybrid Liouvillian framework, we also examined how quantum jumps
affect both chiral transfer and entanglement generation and clarified the
operational meaning of the interpolation parameter $q$ between conditional and
unconditional dynamics. Our results identify an intermediate operating window
that balances non-Hermitian adiabaticity, state-transfer fidelity, and
post-selection cost.  We stress that the present work is not a direct
transposition of known two-level non-Hermitian dynamics to another basis, since
three structural features distinguish the even-parity setting. First,
controlling $|\Phi^{\pm}\rangle$ requires maintaining coherence between the
zero- and two-excitation sectors, so that the EP must be engineered inside a
parity-protected two-photon manifold by parametric two-photon driving and
engineered two-photon loss, which are resources qualitatively different from the
single-excitation-conserving dynamics of the odd-parity scheme. Second, the
engineered jump operator $L \propto \tau_-$ acts entirely within this manifold,
so that quantum jumps degrade but never depopulate the target subspace; this
structural protection is why a sizable chiral contrast ($\Delta\mathcal{F}
\approx 0.77$) and concurrence ($\mathcal{C} = 0.827$) survive even in the fully
unconditional limit $q=1$, and why the scheme responds more mildly to strong
parasitic loss than its odd-parity counterpart (Table~\ref{tab:comparison}).
Third, a single EP-encircling loop serves two purposes, chiral Bell-state
transfer and entanglement generation from a separable state, within one
engineered two-photon platform. Our work thus extends dissipative Bell-state
control beyond the single-excitation sector and may pave the way for
experimental realization in platforms such as superconducting circuits.

\section*{Acknowledgments}
This work was supported by the Natural Science Foundation of Chongqing, China
(Grants No. CSTB2025NSCQ-GPX1272, CSTB2024NSCQMSX0736, and
CSTB2025NSCQ-LZX0142), the National Natural Science Foundation of China (Grants
No. 62571074 and No. 12404145), the research foundation of the Institute for
Advanced Sciences of CQUPT (Grant No. E011A2022328), and the China Postdoctoral
Science Foundation (Grant No. 2025M783465).

\section*{Appendix A}

This appendix serves two purposes. First, it verifies that the even-parity
subspace is exactly decoupled under the full Liouvillian dynamics. Second, it
clarifies the dynamical mechanism underlying the chiral state
transfer in terms of time-integrated low-loss branch selection.

\begin{figure}[htbp]
	\centering
	\begin{minipage}[t]{0.98\linewidth}
		\rlap{(a)\hspace{0.5em}}%
		\includegraphics[width=\linewidth]{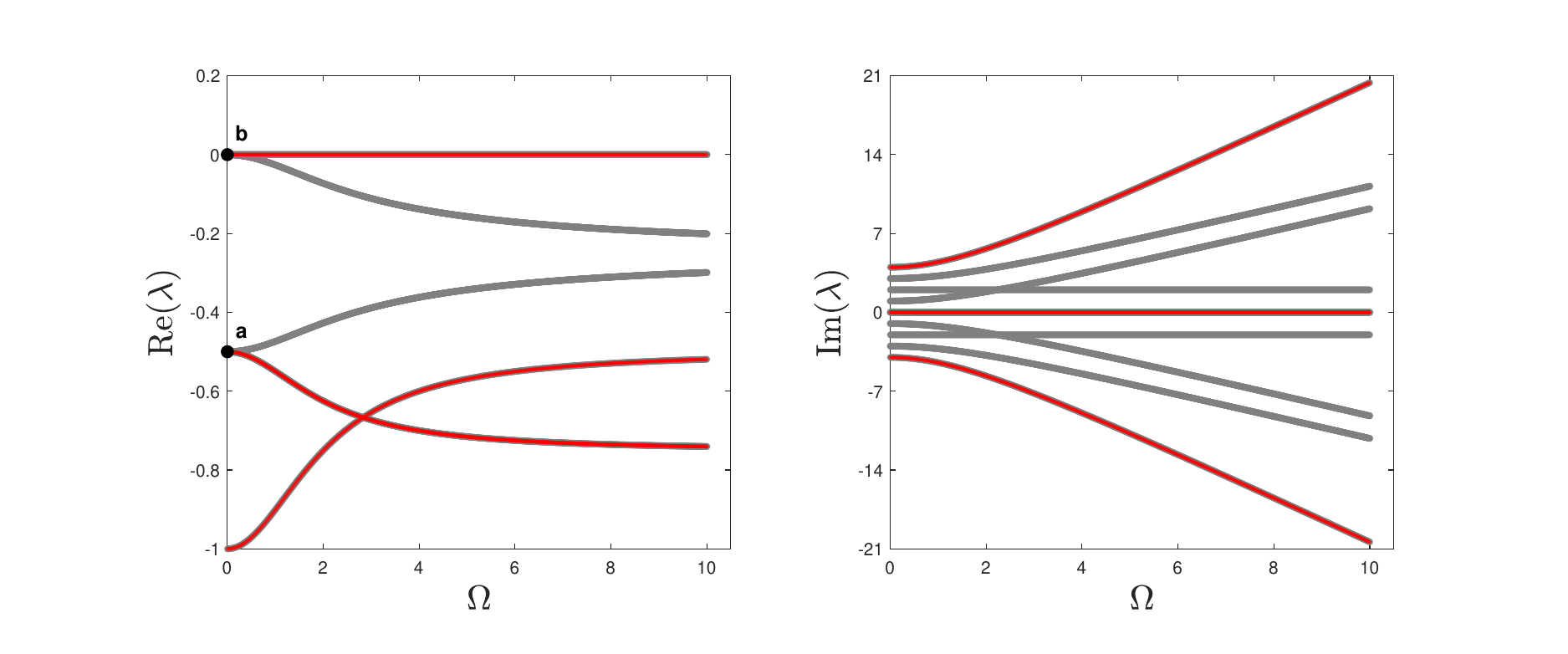}
		\label{subfig:a}
	\end{minipage}
	
	\vspace{0.5em}
	
	\begin{minipage}[t]{0.98\linewidth}
		\rlap{(b)\hspace{0.5em}}%
		\includegraphics[width=\linewidth]{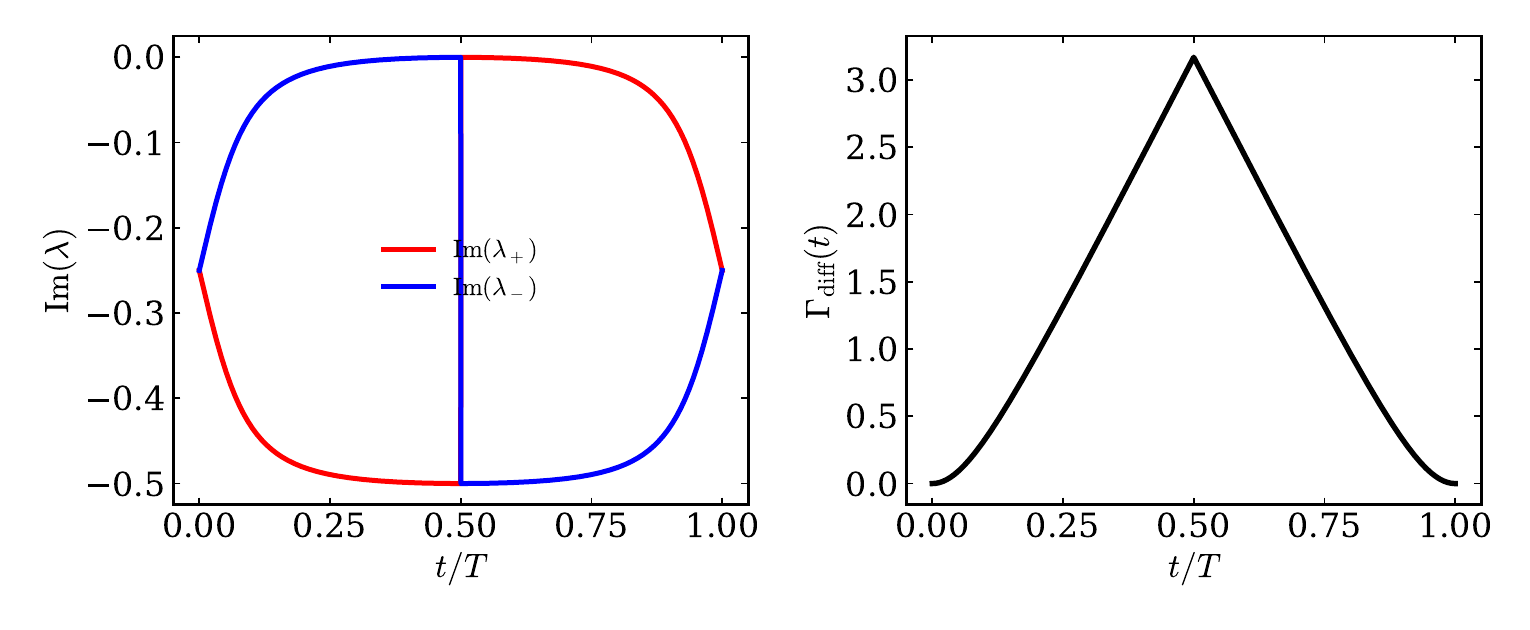}
		\label{subfig:b}
	\end{minipage}
	
	\caption{(a) Comparison between the spectrum of the full Liouvillian
superoperator $\mathcal{L}$ and that of the reduced Liouvillian $\mathcal{L}_1$
restricted to the even-parity subspace. Gray thick lines: all 16 eigenvalues of
$\mathcal{L}$ (left: real parts; right: imaginary parts). Red thin lines: the 4
eigenvalues of $\mathcal{L}_1$. The red spectrum overlaps with a subset of the
gray spectrum, and no avoided level repulsion appears at the marked crossings
(e.g., $a$, $b$). Parameters: $\kappa=1$, $\phi=0$, $\Delta_1=3\kappa$,
$\Delta_2=1\kappa$.
		(b) Left: evolution of the imaginary parts $\operatorname{Im}(\lambda_{\pm})$
of the eigenvalues of the effective Hamiltonian $H_{\text{eff}}$ along the CCW
loop. The shaded regions indicate the instantaneous low-loss branches. Right:
evolution of the cumulative locking factor	$\Gamma_{\mathrm{diff}}(t)=\int_0^t
[\operatorname{Im}(\lambda_{-})-\operatorname{Im}(\lambda_{+})]\,d\tau$.
		The parameters are the same as in Fig.~\ref{fig:F_P_t}.}
	\label{fig:L_L1_gain_loss}
\end{figure}

Figure~\ref{fig:L_L1_gain_loss}(a) compares the eigenvalue spectrum of the full
Liouvillian $\mathcal{L}$ with that of the reduced Liouvillian $\mathcal{L}_1$
acting in the even-parity subspace. Two features are important. First, the four
eigenvalues of $\mathcal{L}_1$ coincide exactly with four of the eigenvalues of
the full Liouvillian, showing that the reduced dynamics is embedded in the full
spectrum. Second, at the marked crossings $a$ and $b$, the eigenvalues
associated with the even-parity sector intersect those of other sectors without
avoided repulsion. Together, these observations are consistent with exact
dynamical decoupling of the even-parity subspace $\{|00\rangle,|11\rangle\}$
from the rest of the Liouvillian dynamics.

This decoupling follows directly from parity symmetry. The two-photon driving
terms in the Hamiltonian, $\sigma_+^{(1)}\sigma_+^{(2)}$ and
$\sigma_-^{(1)}\sigma_-^{(2)}$, change the total excitation number by $\Delta
N=\pm2$, thereby coupling only $|00\rangle$ and $|11\rangle$ and leaving the
odd-parity states $|01\rangle$ and $|10\rangle$ unaffected. The dissipation
operator $L=\sqrt{\kappa}\,\sigma_-^{(1)}\sigma_-^{(2)}$ also changes the
excitation number by $\Delta N=-2$, mapping $|11\rangle$ to $|00\rangle$ without
acting on the odd-parity sector. As a result, the full Liouvillian preserves
parity exactly, and no transitions occur between subspaces of different parity.

In non-Hermitian dynamics, adiabatic following is influenced not only by the
instantaneous eigenvectors but also by the exponential amplification or
suppression factors associated with the imaginary parts of the eigenvalues.
Figure~\ref{fig:L_L1_gain_loss}(b) illustrates this effect for the effective
Hamiltonian $H_{\text{eff}}$ along the CCW encircling path. During the first
half of the loop, $\operatorname{Im}(\lambda_{-})$ remains larger than
$\operatorname{Im}(\lambda_{+})$, which means that the eigenstate associated
with $\lambda_{-}$ experiences a smaller cumulative loss and therefore acquires
a survival advantage. This is a direct manifestation of the standard principle
of non-Hermitian dynamics that the imaginary parts of the eigenvalues govern the
exponential decay of each instantaneous
eigenstate~\cite{uzdinObservabilityAsymmetryAdiabatic2011c,%
dopplerDynamicallyEncirclingExceptional2016e}. The branch with lower loss
accumulates a survival advantage over time, which is quantified by the
cumulative locking factor $\Gamma_{\mathrm{diff}}(t)$ defined below.

To quantify this imbalance, we define the cumulative locking factor
\begin{equation}
	\Gamma_{\mathrm{diff}}(t) = \int_0^t \left[ \operatorname{Im}(\lambda_{-}) -
\operatorname{Im}(\lambda_{+}) \right] \, d\tau .
\end{equation}
Except at the beginning and end of the cycle, $\Gamma_{\mathrm{diff}}(t)$
remains positive throughout the evolution. It first increases, then decreases,
and reaches its maximum at $t=T/2$. This behavior indicates that the low-loss
branch accumulates a substantial integrated advantage during the first half of
the loop, so that the evolving state becomes preferentially locked onto the
branch associated with $\lambda_{-}$.

The resulting chiral state transfer can therefore be understood as the combined
effect of two ingredients. The first is branch exchange associated with EP
encircling, which determines the geometric structure of the evolution. The
second is time-integrated low-loss branch selection, which biases the dynamics
toward the less dissipative branch. In this sense, the ``midway locking''
discussed in the main text is a convenient description of branch selection
induced by cumulative loss asymmetry, consistent with non-Hermitian state
conversion in dynamically encircling
EPs~\cite{uzdinObservabilityAsymmetryAdiabatic2011c,%
dopplerDynamicallyEncirclingExceptional2016e}. Under the CCW loop, this
mechanism ultimately locks the system onto the eigenstate $ |\Phi_{-}\rangle(T)
$, in agreement with the projection dynamics shown in Fig.~\ref{fig:F_P_t}(b).
Note that the present scheme differs from the odd-parity chiral transfer of
Ref.~\cite{khandelwalChiralBellStateTransfer2024} at the structural rather than
the mechanistic level. In both cases the transfer is governed by branch
selection near a second-order exceptional point of the effective non-Hermitian
Hamiltonian, but here the exceptional point resides in a parity-protected
two-dimensional manifold whose engineered jump operator $L \propto \tau_-$ acts
entirely within the manifold, whereas in the single-excitation scheme the
engineered gain and loss jumps connect the working manifold to the zero- and
two-excitation sectors (see Table~\ref{tab:comparison}).

\end{document}